\documentstyle[aps,12pt]{revtex}
\input epsf
\begin{document}

\def\la{{\langle}}
\def\ra{{\rangle}}

\def\be{\begin{equation}}
\def\bea{\begin{eqnarray}}
\def\ee{\end{equation}}
\def\eea{\end{eqnarray}}

\title{Efficiency of DNA replication in the Polymerase Chain Reaction\\
\protect{\small (polymerization reaction/branching processes/kinetic 
model/quantitative PCR)}}
\author{Gustavo Stolovitzky and  Guillermo Cecchi }
\address{Center for Studies in Physics and Biology, The Rockefeller University, \\
1230 York Avenue, New York NY 10021\\
}
\maketitle
\protect \vspace{15cm}

\noindent Classification: Biophysics.

\noindent Abbreviation: pdf, probability density function.
\newpage

\begin{abstract}
{\bf ABSTRACT}
A detailed quantitative kinetic model for the Polymerase Chain Reaction 
(PCR) is developed,
which allows us to predict the probability of replication of a DNA 
molecule
in terms of the physical parameters involved in the problem. 
The important issue of the determination of the number of PCR cycles
during which this probability can be considered to be a constant
is solved within the framework of the model. New phenomena of 
multi-modality 
and scaling behavior in the distribution of the number of molecules after
a given number of PCR cycles are presented. The relevance of
the model for quantitative PCR is discussed, and a novel quantitative
PCR technique is proposed.
\end{abstract}
\newpage
\narrowtext

\baselineskip .29truein

\section{INTRODUCTION}
The Polymerase Chain Reaction (PCR) is one of the most widely used
techniques in modern molecular biology. It was devised \cite{MULLIS} as
a method for amplifying specific DNA sequences (targets), and the scope
of its applications stretches from medicine \cite{MEDICINE}, through
{\sl in vitro} evolution \cite{VITRO}, to molecular computers
\cite{ADLEMAN,KAPLAN}. In spite of its ubiquity in biology, theoretical
discussions of PCR are rare. Although kinetic models of the
enzyme-mediated polymerization of single-stranded DNA have been reported
\cite{CAPSON,PATEL,BENKOVIC}, none of them were applied to model
PCR, and only recently a 
treatment of the rate of mutations
arising in PCR has been considered \cite{SUN,WEISS}.

The main object of our study is the probability that one molecule will
be replicated in one PCR cycle, the so-called efficiency $p$.  
In Section II we present a
detailed kinetic model of the polymerization and find $p$ as a
function of the physical parameters of the problem. This allows us to
discuss the range of validity of the assumption of constant
probability of replication, on which statistical
considerations have been based \cite{SUN,WEISS}. Within that range, we
apply the theory of branching processes in Section III,
to show the existence of new
phenomena:  the probability density function (pdf) of the number of
molecules after a given number of cycles of PCR
displays scaling behavior, and under some conditions, this pdf is 
multi-modal. In Section IV a novel method for quantitative PCR
is presented, based on the statistical considerations of the previous
sections. In Section V we summarize our work.

One cycle of PCR consists of three
steps. (For a more detailed account of the PCR technique see, e.g., \cite{PCR}.) 
In the {\sl denaturing} step, the two strands of the parent DNA
molecule in solution are separated into single-stranded (ss)
templates by rising the temperature to about ${95^{\circ}} C$ to
disrupt the hydrogen bonds. In the {\sl annealing} step, the solution is
cooled down to approximately ${50^{\circ}} C$ to allow the
{\sl primers}, present in a high concentration, to hybridize with the
ss DNA. The primers are two (different) 20 to 30 nucleotides long
molecules which are Watson-Crick complementary to the 3' flanking
extreme of the templates. Once the primer-template heteroduplex is
formed, it acts as the {\sl initiation complex} for the {\sl
DNA-polymerase}\footnote{Polymerases are interesting 
pieces of machinery 
 \cite{POLY}. 
%They stand in the very core of life, as 
They are responsible for the duplication of
genetic information (DNA-polymerases) and its transcription into RNA
(RNA-polymerases).} to recognize and bind to. This step is
crucial for the specificity of the amplification: only those molecules
that have sequences complementary to the primers will be amplified. The
last step is a polymerization reaction, in which the solution is heated
to ${72^{\circ}} C$, the optimal working temperature for {\sl Thermus
Aquaticus} DNA
polymerase. This enzyme catalyzes the binding of complementary
nucleotides to the template, in the direction that goes from the primer
to the other extreme\footnote{The DNA is a $polar$ molecule, and the polymerase can
only attach new nucleotides to the 3' end of the molecule that is being
extended.}. Notice that if this polymerization
proceeded to its end, at the end of the third step we would have twice
as many DNA molecules as we had at the beginning of Step 1. These three
steps constitute one cycle of the PCR, which is usually 30 seconds to
2 minutes long. The cycles are repeated a number
of times (typically 30) by varying the temperature in the solution, in
such a way that the DNA molecules that were synthesized in a given
cycle are used as templates in the following one. In this way one gets
an extremely efficient amplification mechanism for DNA.

\section{KINETIC MODEL}
We will represent the last two steps of a typical cycle of PCR 
by means of a kinetic model. Our species will be the primers
($pr$, of length $L_p$ nucleotides), the ss DNA ($ss$, consisting of
$L_p +N$ nucleotides), the heteroduplexes ($h_i$, formed by one complete 
$ss$ and the partially
assembled complementary strand consisting of the primer and the next $i$ 
nucleotides), the nucleotides ($n$, which will be considered identical), the
polymerase ($q$) and the heteroduplexes $h_i$ with the polymerase 
attached to them ($qh_i$). Denoting by $\kappa_{2j-1}$ and $\kappa_{2j}$
the forward and backward chemical reaction rates, the chemical 
equations are
\begin{eqnarray}
& &\mbox{Step 2}\hspace{.15cm} \{\hspace{.65cm}ss + 
pr~~ \stackrel{\kappa_{1,2}}{\rightleftharpoons}~~ h_{0} \nonumber  \\
& &\mbox{Step 3} \left \{\begin{array}{l}\left \{ \begin{array}{c}
h_{i} + q~~ \stackrel{\kappa_{3,4}}{\rightleftharpoons}~~ qh_{i} \nonumber \\
\hspace{.15cm}qh_{i} + n~~ \stackrel{\kappa_{5,6}}{\rightleftharpoons}~~ qh_{i+1} 
\nonumber 
\\
\end{array} \right \} \hspace{0.5cm} 0 \le i \le N-1
 \nonumber 
 \\
\hspace{.65cm}h_{N} + q~~ \stackrel{\kappa_{7,8}}{\rightleftharpoons}~~ 
qh_{N}. \end{array} \right. \\  \label{chem_reac}
\end{eqnarray}
Other recognizable species might be present
in the chain reaction. This can occur because of
substitutions, additions or deletions
of nucleotides by the polymerase, or because of the presence of 
sequence-dependent
structures. These will not be taken into account in our model, for
the sake of simplicity. 
%
% Moreover, many DNA-polymerases have exonucleolytic activity,
% such as editing of mismatches. However, since {\sl Thermus Aquaticus}
% is the most common polymerase for PCR applications \cite{TINDAL,KIM}, 
% the exonucleolytic activity is not included in our model.
%
Assuming that the effects of inhomogeneities in density and
temperature are irrelevant, it is well known that Eqs.~(\ref{chem_reac}) lead to
a corresponding system of first order nonlinear differential
equations for the concentrations of the different species 
as functions of time, which we are not going
to write here (see, for example, \cite{VANKAMPEN}). In the above reactions, one should
assign a given duration to Step 2 and another to Step 3. 
For the sake of simplicity, however, 
we shall consider both Step 2 and Step 3 as running simultaneously in the
simulations to be presented below. This is a mild simplification which does not
alter the conclusions to be drawn. 
 
The definition of the efficiency $p$ implies that it is simply
the ratio between the number of $ss$ molecules
that were completely replicated at the end of a given cycle and the
initial number of $ss$ molecules in that cycle:
\be
p(t) = \frac{[h_N](t)+[qh_N](t)}{[ss](0)}
\label{p}
\ee
Figure 1 shows plots of the probability of replication $p$ as a function
of time $t$ (which is to be interpreted as the duration of Step 3 in a 
typical PCR cycle), for different polymerization lengths $N$. 
%The values
%of the chemical reaction constants $\kappa$ and the initial 
%conditions of the simulation
%are detailed in the caption to Fig. 1. 
Since to the best of our knowledge the chemical reaction constants
$\kappa$'s have not been measured for Taq polymerase, we have 
assumed some values for these constants to exemplify the principal
characteristics of our model. It should be stressed, however, that the
equivalent to some of these constants have been measured for other
polymerases such as T4 polymerase \cite{CAPSON}, T7 polymerase 
\cite{PATEL} and for DNA polymerase I (Klenow fragment) \cite{BENKOVIC}.
The values
of the chemical reaction constants $\kappa$ and the initial 
conditions used in the simulation of Eq.~(\ref{chem_reac}) 
are detailed in the caption to Fig. 1. 
The main features of the curves in Fig. 1 can be 
quantitatively understood. It can be observed that the larger $N$, the flatter
the behavior at small times. Indeed it can be shown from the dynamic equations
that $p \sim t^{(2N+1)/3}$ for $t$ sufficiently small. The time at which $p$ has
reached about half its asymptotic value, as well as the width of the rise-time
 can be estimated from a further simplification of our model. Assuming that
the time constants associated with the backwards reactions in Eq.~(\ref{chem_reac}) 
are large enough, and that the concentration of primers, polymerase
and nucleotides are sufficiently large that their relative concentrations can
 be considered as constants (or more precisely as slowly varying parameters) 
during the process, we can rewrite the reaction as
\be
ss \stackrel{\stackrel{pr}{\downarrow}\kappa_{1}}{\rightarrow} h_{0}
\stackrel{\stackrel{q}{\downarrow}\kappa_{3}}{\rightarrow}qh_0
\stackrel{\stackrel{n}{\downarrow}\kappa_{5}}{\rightarrow}qh_1 \dots
\stackrel{\stackrel{n}{\downarrow}\kappa_{5}}{\rightarrow}qh_N.
\label{forward}
\ee
The time $\tau$ needed for this reaction to be completed is simply the 
sum of the times corresponding to
each link of the chain, $\tau=\tau_{\kappa_1}+\tau_{\kappa_3}
+\tau_{\kappa_5,1}+...+\tau_{\kappa_5,N}$, where $\tau_{\kappa_1}$ and
$\tau_{\kappa_3}$ are the times associated with the first two reactions in 
Eq.~(\ref{forward}), and 
$\tau_{\kappa_5,j}$ is the time associated with the reaction 
$qh_{j-1}
\stackrel{\stackrel{n}{\downarrow}\kappa_{5}}{\rightarrow}qh_j$. 
These $\tau$'s are
independent, exponentially distributed
random variables, whose mean values are
$\la \tau_{\kappa_1} \ra = (\kappa_1 [pr])^{-1}$, 
$\la \tau_{\kappa_3} \ra = (\kappa_3 [q])^{-1}$, and
$\la \tau_{\kappa_5,j} \ra = (\kappa_5 [n])^{-1}$. Therefore, 
it can be readily seen that the mean and the standard deviation of $\tau$ are
\bea
\la \tau \ra  & = &  \frac{1}{\kappa_1 [pr]} +  \frac{1}{\kappa_3 [q]}
+ N\frac{1}{\kappa_5 [n]}, \label{meantau}\\
\sigma_{\tau} & = & \left( \frac{1}{\kappa_1^2 [pr]^2}+  
 \frac{1}{\kappa_3^2 [q]^2}
+ N  \frac{1}{\kappa_5^2 [n]^2}\right) ^{1/2},\label{vartau}
\eea
where we used that the variance of a exponentially distributed random variable is the 
square of its mean. $\la \tau \ra$ and $\sigma_{\tau}$ can be used as estimates of 
mean rise-time and the rise-time width about the mean for the complete reaction. 
These estimates are shown in Fig. 1. The abscissa of the solid square on each curve 
corresponds to the value predicted by Eq.~(\ref{meantau}), and the arrow heads indicate
the values of $\la \tau \ra \pm \sigma_{\tau}$. It can be safely concluded that 
Eqs.~(\ref{meantau}) and (\ref{vartau}), computed from the simplified
chain reactions of Eq.~(\ref{forward}), are good estimates of the mean rise-time 
and the rise-time width corresponding to the full set of reactions. 

The last important feature to be extracted from Fig. 1 is the tendency of $p(t)$ towards
an asymptote $p_{\infty}$, which corresponds to the equilibrium of the chemical
system. This value is of importance in PCR, and thus
it is worth computing it in terms of the parameters of our model. The detailed 
balance equilibrium conditions for the reactions of Eq.~(\ref{chem_reac}) demand
that $[ss]_{eq}/[h_0]_{eq}=\kappa_2/(\kappa_1 [pr]_{eq}) \equiv \alpha_1$, 
$[h_i]_{eq}/[qh_i]_{eq}=\kappa_4/(\kappa_3 [q]_{eq}) \equiv \alpha_3$ (for
$0 \le i \le N-1)$, 
$[qh_i]_{eq}/([qh_{i+1}]_{eq})=\kappa_6/(\kappa_5 [n]_{eq}) \equiv \alpha_5$ 
(for $0 \le i \le N-1$) and
$[h_N]_{eq}/[qh_{N}]_{eq}=\kappa_8/(\kappa_7 [q]_{eq}) \equiv \alpha_7$.
On using Eq.~(\ref{p}) and the conservation relation
 $[ss](t)+\sum_{i=0}^{N} [h_i](t) + [qh_i](t)=[ss](0)$, one obtains that
\be
p_{\infty} = \frac{1+\alpha_7}{1 + \alpha_7 + \alpha_1 \alpha_3 \alpha_5^N +
\frac{\alpha_3+1}{\alpha_5-1}(\alpha_5^{N+1} - \alpha_5)}.
\label{p_inf}
\ee
For the purpose of computing $p_{\infty}$ one should know the values of $[pr]_{eq}$,
$[n]_{eq}$ and $[q]_{eq}$. As an approximation to these values one can use the 
initial values of these species at the beginning of the cycle. This approximation will
be excellent if this initial concentrations are sufficiently large. The values of 
$p_{\infty}$ (computed under this approximation) corresponding to the conditions 
of the simulations of Fig. 1 are $0.87$ for $N=1$ and $0.85$ for $N=10$ and $N=45$, in
perfect agreement with the complete simulation.
It is interesting to notice that from direct
measurements of $p(t)$ a wealth of 
information on the rate constants involved in the 
polymerization reactions can be inferred using 
Eqs.~(\ref{meantau})-(\ref{p_inf}).  

Of utmost importance in applications of PCR is the number
of cycles of PCR during which the amplifying process is exponential. As will
be discussed later on, the mean number of molecules $\la N_{k+1} \ra$
at cycle $k+1$ is related to the mean number of molecules $\la N_{k} \ra$ 
at cycle $k$ by the relation $\la N_{k+1} \ra = (1+p_k) \la N_{k} \ra$, where
$p_k$ is the efficiency during the $k$-$th$ cycle.
Therefore the rate of growth will be exponential only when $p_k$ is independent
of $k$. During how many cycles can the system maintain $p_k$ constant? The answer 
can be 
found if we think that during these cycles, both the concentration of primers and 
nucleotides will also be decreasing exponentially, and therefore their concentration
at cycle $k$ will be $[pr]_k=[pr]_0-(1+p)^k [ss]_0$ and  $[n]_k=[n]_0-(1+p)^k 
N [ss]_0$.
The mean rise-time and rise-time width for $p$ at cycle $k$, $\la \tau \ra_k$ and
$\sigma_{\tau,k}$,
 will be given by Eqs.~(\ref{meantau}) and (\ref{vartau}), with
$[pr]$ and $[n]$ replaced by $[pr]_k$ and $[n]_k$ respectively. If the time for the 
reaction is $t$, then the maximum number of cycles $\nu$ during which $p_k$ can be 
considered constant will be given, to a first approximation, by the $\nu$ that 
verifies
that $\la \tau \ra_{\nu}+\sigma_{\tau,\nu}= t$. This imposes an equation for $\nu$ 
that 
can be solved numerically. An approximation to this solution is 
\bea
\nu =  \min & &\left \{ \log_{1+p_{\infty}} \left ( \frac{[n]_0}{[ss]_0 N} -
\frac{1}{[ss]_0 \kappa_5 (t - \kappa_3^{-1} [q]^{-1})} \right ) \right.;\nonumber \\
& & \left .\log_{1+p_{\infty}} \left ( \frac{[pr]_0}{[ss]_0} -
\frac{1}{[ss]_0 \kappa_1 (t - \kappa_3^{-1} [q]^{-1})} \right ) \right \}
\label{nu}
\eea
where $log_b$ indicates logarithm to the base $b$. 
Notice that as $t$ becomes larger, the value of $\nu$ predicted in
Eq.~(\ref{nu}) tends to a constant independent of $t$, given by the number of cycles 
that it takes to deplete the solution of nucleotides or primers, 
whichever is exhausted first. Although
it might be unrealistic for the conditions used in molecular biology, 
it is interesting
to notice that if the nucleotides are the first species to be exhausted, 
then most of the
heteroduplexes will cease to polymerize before reaching the end, with the outcome
that there will hardly be any complete double helix formed: in this case 
the net amplification factor will be close to zero.
Figure 2 shows the efficiency
$p_k$ at cycle $k$ as a function of the number of cycles, for different
times of polymerization and $N=45$ (the other parameters are as in Fig. 1), 
obtained by the integration of the reactions of 
Eq.~(\ref{chem_reac}). In this simulation we concatenated cycles assuming
a perfect melting step, which was done by hand by setting $[ss]_{k+1}(0)$ in the cycle
$k+1$ equal to $[ss]_k(0)+[h_N]_k(t)+[qh_N]_k(t)$ of the previous cycle and
$[h_i]_{k+1}(0)=[qh_i]_{k+1}(0)=0$, ($0 \le i \le N$). The dynamics of $pr$ and $n$, on the other hand, was followed
exactly. 
It is clear from Fig. 2 that there is a regime for which $p_k$
is roughly constant, and that the extent of this regime tends to decrease
with $t$. The values of $\nu$ predicted by the condition 
$\la \tau \ra_{\nu}+\sigma_{\tau,\nu}= t$ are $13$ for $t=0.8$ $sec$ 
and $15$ for $t=2.0$ $sec$ 
[slightly overestimated by Eq.~(\ref{nu}), whose integer part yields 
$14$ for $t=0.8$ $sec$ and 
$15$ for $t=2.0$ $sec$], in rough agreement with the values of about $12$ and $14$ respectively
obtained from Fig. 2. 

At this point a few important considerations are in order.
The fraction $1-p$ of molecules whose replication was incomplete will
give rise to incomplete complementary single strands. Only when
these incomplete replicas are close to completion will they be able to bind 
a primer in the next cycle, and thus be replicated. Therefore the efficiency
$p$ defined in Eq.~(\ref{p}) is an underestimation, since $h_{N-1}$, $h_{N-2}$,
$\dots$, $h_{N-j}$ as well as $qh_{N-1}$, $qh_{N-2}$,
$\dots$, $qh_{N-j}$ (for some $j<L_p$, where $L_p$ is the length of the 
primers), will
be part of the pool of templates in subsequent cycles. However, the dominant 
process
will be the replication of the complete strand, which justifies the computation 
of
$p$ as in Eq.~(\ref{p}). There is another issue that needs some discussion.
All the complementary strands arising from both complete and incomplete
replication of a template
can anneal to that template in subsequent cycles, and therefore can act 
effectively as primers. Strictly speaking, at any 
given
cycle $k \ge 1$ there will be a pool of primers of different lengths. An estimate 
of the concentration of ``primers'' arising from incomplete replication 
at cycle $k$ 
is $\frac{1-p}{p}(1+p)^k [ss](0)$. This amount is always smaller than the 
concentration $(1+p)^k [ss](0)$ of completely replicated single strands 
which act also as potential ``primers''.
As long as the concentration of incomplete replicas remain much smaller
than the concentration of primers $[pr]$ , Eqs.~(\ref{chem_reac}) 
will constitute
a good approximation to the PCR process. Recall now that $\nu$ [see
Eq.~(\ref{nu})] is equal or smaller than the number of cycles required for the
concentration of primers $[pr]$ to match the concentration of completely 
replicated
single stranded molecules. It follows that the  approximation given by 
Eqs.~(\ref{chem_reac}) will break down
only after the number of PCR cycles is bigger than $\nu$
and therefore our basic conclusions, contained in Eqs.~(\ref{meantau})-(\ref{nu}),
are not altered.

\section {STATISTICAL ANALYSIS}
As seen above, the efficiency $p$ can be assumed to 
be constant for a number of cycles of PCR. The 
statistics of PCR can be readily computed under
this assumption.
The basic element in the analysis is the recursive relation
that links the number of replicates after cycle number $n+1$,
 $N_{n+1}$ in terms of $N_n$,
\be
N_{n+1}= N_n + B(N_n;p)
\label{recursion}
\ee
where $B(N_n;p)$ is a random variable whose 
distribution is binomial with parameters $N_n$ and $p$. The basis for 
this relation is that at the $n+1$-$th$ cycle, there will be not only
the $N_n$
molecules that were present at the previous cycle, but also the number
of successful replication after $N_n$ Bernoulli trials 
\cite{FELLER}
each one with probability $p$ of success. The number of
molecules in the initial sample will be denoted by $M_0$.

The first moments of $N_n$ can be easily 
computed from Eq.~(\ref{recursion}):
\bea
& &\mu_n \equiv  \langle N_n \rangle =  M_0 (1+p)^n,\label{mean}\\
& &\sigma^2_{n} \equiv \langle ( N_n - \mu_n)^2 \rangle 
 =  M_0 \frac{1-p}{1+p}\left [ (1+p)^{2n} - (1+p)^n \right ].\nonumber\\
  \label{variance}
\eea
Furthermore, using the theory of branching processes \cite{HARRIS,NEY},
a recursive relation between $P_{n}^{M_0}(k)$ (the probability that
there are $k$ molecules at cycle $n$, having started with $M_0$ of them) and 
$P_{n-1}^{M_0}(k)$ can be obtained
\be
P_{n}^{M_0}(k) = \sum_{j=[\frac{k}{2}]}^{j_{max}}
\left ( \begin{array}{c} j \\ k-j \end{array} \right ) 
p^{k-j} (1-p)^{2j-k}  P_{n-1}^{M_0}(j),
\ee
(where $j_{max}=\min\{M_0 2^{n-1}, k\}$, and 
$[\frac{k}{2}]$ denotes the integer part of $\frac{k}{2}$), 
which when supplemented with the initial condition $P_{0}^{M_0}(k)=
\delta_{k,M_0}$ allows us to compute $P_{n}^{M_0}(k)$ for any
$n$. Figure 3 shows the form of these probability functions for $n=10$
with   
$M_0=1$ in Fig. 3(a), and $M_0=50$ in Fig. 3(b), and different values of $p$. A
remarkable resonance-like behavior can be observed in the curve
corresponding to $p=0.9$ and $M_0=1$ [wavy curve in  
Fig. 3(a)]. This phenomenon originates in the discrete nature of
the process: if at the first cycle the system fails in replicating
the only original template, then the subsequent growth of the population
will be as if there were nine cycles instead of ten. The
other peaks correspond to the failure in replication in the 
first two cycles, three cycles, etc. This trait is characteristic of
values of $p$ between say $0.8$ to $1$. For smaller values of $p$
the function looks smoother. A common feature of the curves in Fig.~3(a)
is the existence of a power law regime in the region of
small $N_n$, whose origin will be discussed later on.
The behavior of the curves with $M_0=50$ is simpler: they are basically
Gaussian curves, with a mean that increases with $p$ and a variance that
first increases and then decreases with $p$ [see Eqs.~(\ref{mean}) and 
(\ref{variance})].

In order to understand the features described above, it is convenient to
use the formalism of generating functions \cite{FELLER}. The generating function
of $P_n^{M_0}(N_n)$ is simply $g_{n,M_0}(s)=\la s^{N_n} \ra$. Using Eq.~(\ref{recursion})
it is clear that $g_{1,1}=(1-p)s+ps^2$. It can be shown \cite{HARRIS} that for a branching 
process
\bea
g_{n,M_0}(s)  =  g_{n-1,M_0}[g_{1,1}(s)]  =
 \dots = 
 \left [ g_{1,1}^{(n)}(s) \right ]^{M_0}, \label{composition}
\eea
where we have denoted by $g_{1,1}^{(n)}(s)$ the $n$-$th$ composition of 
$g_{1,1}(s)$ with itself, and used that $g_{0,M_0}(s)=s^{M_0}$ in the last
equality.
To proceed, we use the formalism of characteristic functions. 
The characteristic
function $\phi_{n,M_0}(\omega)$ of the distribution of $N_n$ having started with 
$M_0$ molecules, which is 
by definition the Fourier transform of $P_n^{M_0}(N_n)$ \cite{FELLER}, 
is  simply $\phi_{n,M_0}(\omega)=g_{n,M_0}(e^{i\omega})$. In terms
of the characteristic functions, Eq.~(\ref{composition}) implies
that 
\be
\phi_{n,M_0}(\omega)= \left[ \phi_{n,1}(\omega) \right]^{M_0}.
\label{charfunc}
\ee
The characteristic function
of the sum of $M_0$ {\em independent} random variables is simply the 
product of the characteristic functions of each of them. Therefore, 
the physical interpretation of the last equation is that the 
amplification cascades
produced by each of the $M_0$ original molecules proceed independently,
without interaction. From this observation and
the central limit theorem it follows that as the number
of molecules $M_0$ becomes larger, the distribution of $N_{n}$
tends to a Gaussian. 
This explains the observed features of the pdfs of Fig. 3(b).

The behavior of the pdfs for finite 
$M_0$ in the limit of $n \rightarrow \infty$ is a little bit more 
interesting. In fact, it is clear from Eq.~(\ref{charfunc})
that it suffices to study the case $M_0=1$, which we do next.
We should stress that our study of the asymptotically large
$n$ regime does not aim at understanding the behavior
of PCR when infinitely many cycles are performed. In fact we have shown
in the previous Section that the efficiency can be considered as constant
only for a finite number of cycles. Rather, the reason for studying
this asymptotic regime is that the convergence of the finite $n$ case
to the $n \rightarrow \infty$ case is fast enough that many of the
features arising for finite $n$ are well explained by the study of asymptotically
large $n$, most notably the power law behavior of the low $N$ regime of Fig. 3(a).
It follows from Eq.~(\ref{composition}) that 
$g_{n,1}(s)=g_{1,1}[g_{n-1,1}(s)]$, which in terms of the characteristic
functions and of the explicit  expression for $g_{1,1}(s)$ becomes
\be
\phi_{n,1}(\omega)=(1-p) \phi_{n-1,1}(\omega) + p [\phi_{n-1,1}(\omega)]^2.
\label{map0}
\ee
Given that
we are going to consider the limit of $n \rightarrow \infty$ and
$\langle N_n \rangle=(1+p)^n$ 
[see Eq.~(\ref{mean})] diverges in this limit, it is
convenient to use the random variable 
$\tilde{N_n}=N_n/\langle N_n  \rangle$. Denote by $\theta_{n,1}(\omega)$
its characteristic function. It is easy to show that $\theta_{n,1}(\omega)=
\phi_{n,1}(\frac{\omega}{(1+p)^n})$, which on using Eq.~(\ref{map0}) yields
\be
\theta_{n,1}(\omega) = (1-p) \theta_{n-1,1}(\frac{\omega}{1+p}) +
 p [\theta_{n-1,1}(\frac{\omega}{1+p})]^2.
\label{map}
\ee
Notice that Eq.~(\ref{map}) can be thought of as a dynamical system,
that maps the point $z_n$ to $z_{n+1}\equiv f(z_n)=(1-p) z_n + p z_n^2 $. 
The function
$\theta_{n,1}(\omega)$ ($-\infty \le \omega \le \infty$) 
 parametrizes a curve in the complex plane. In fact, the initial
condition $M_0=1$ determines that $\theta_{0,1}(\omega)=e^{i\omega}$, which 
parametrizes the unit circle $\zeta_0$. Subsequent
applications of 
the map $f(z)$ to $\zeta_0$ produces the new curves $\zeta_1$, $\zeta_2$,
$\dots$, which are parameterized respectively by $\theta_{1,1}(\omega)$,
$\theta_{2,1}(\omega)$, $\dots$. The study of the limiting behavior of the
pdf of $\tilde{N_n}$ is thus associated with the study of the invariant
curves of the map $f$. Notice that the map $f$ has only two fixed 
points, one
at $z=0$ (stable) and one at $z=1$ (unstable). Upon iteration, all 
the infinitesimally 
small straight lines with slope $\lambda$ passing through 
the repelling point $z=1$ 
will generate a curve ${\it C}_{\lambda}$ which is invariant 
under $f$, that is $f({\it C}_{\lambda})={\it C}_{\lambda}$. 
On the other hand, for
any $z\ne1$ such that $|z|\le 1$, $|f(z)|<|z|$. Therefore the dynamics
of this map brings all the points of $\zeta_0$  (except for $z=1$) 
to the origin. In the neighborhood of $z=1$, $\zeta_0$ is locally a straight
line with slope $\lambda=\infty$, which upon evolution will become the 
invariant manifold ${\it C}_{\infty}$. It follows that $\zeta_{\infty}$
coincides with ${\it C}_{\infty}$, and $\theta_{\infty,1}$ parameterizes
the invariant manifold of the map $f$, that crosses $z=1$ parallel to the imaginary
axis. Figure 4(a) shows half the invariant manifold ${\it C}_{\infty}$ 
corresponding to $p=0.9$ (the other half is its complex conjugate), and 
on the same plot the imaginary part $vs$ the real part of $\theta_{15,1}(\omega)$
(for positive $\omega$).
To the level of resolution of the figure no departures between the two 
curves are observed, meaning that the pdf of the number of 
molecules at $15$ PCR cycles is well approximated by the limiting pdf.

This dynamical-system way of looking at the characteristic function of $\tilde{N_n}$
is very useful to understand the power law behavior of the pdfs of $N_n$.
The argument goes as follows. Close to $z=0$ (or equivalently, for large values 
of $\omega$) the quadratic terms in
Eq.~(\ref{map}) can be neglected, and the resulting approximate relation,
$\theta_{\infty,1}(\omega)=(1-p) \theta_{\infty,1}(\omega/(1+p))$ accepts as a solution
the ansatz $\theta_{\infty,1}(\omega) \approx A(\ln \omega) 
\omega^{\frac{\ln (1-p)}{\ln (1+p)}}$, where $A(x)$ is in principle any periodic
function with period $\ln(1+p)$. The large $\omega$ behavior of the characteristic
function $\theta_{\infty,1}(\omega)$ is then a power law, with logarithmically 
periodic modulations. That this is so is shown in Fig. 4(b), where we have 
plotted the absolute value of $\theta_{15,1}(\omega)$
for $p=0.9$. The power law corresponding to the predicted scaling exponent of 
$\frac{\ln (1-p)}{\ln (1+p)}$ is shown as the straight line close to the
curve in the log-log plots of Fig. 4(b). The implication of this results
for the pdfs can be readily drawn. Recalling that the characteristic function 
and pdf are related through a Fourier transform, and that the
Fourier transform of $|\omega|^{\alpha}$ (with an appropriate infrared cut-off) 
scales as $x^{-\alpha -1}$, we conclude that the pdf of $\tilde{N_n}$
should exhibit a scaling of the form $P_{\infty}^{1}(\tilde{N_n}) \sim 
\tilde{N_n}^{-\frac{\ln (1-p)}{\ln (1+p)}-1}$. This scaling law is 
shown as the straight lines close to the curves plotted in log-log
scale in Fig. 3(a).

In the following Section we apply some of the results presented so far
to the problem of quantitative PCR.  

\section{QUANTITATIVE PCR}
Although PCR is used mainly in a qualitative fashion, its
potential for becoming an important tool in nucleic acid
quantification  in general \cite{FERRE}, and in medical research in particular
\cite{CLEMENTI} has become clear in recent years. By quantitative PCR
one means the use of the PCR to measure an unknown initial number of molecules
 $M_0$. A few techniques have been developed to that effect in the past, but the
most
widespread is probably the so-called competitive PCR (see, e.g.,
\cite{GUILLILAND}). In this technique, the target, whose initial
concentration is unknown, is amplified simultaneously
with a standard, which is flanked by the same primers as the target and whose
initial concentration is known. The standard should have a length different
from that of the target, so that both can be resolved in an electrophoretic
gel. The basic idea in competitive PCR is that {\em if the efficiencies of
replication of the target and the standard are the same} then the
ratio of the concentration of target to that of the standard is constant
in the reaction. Measuring that ratio at cycle $n$ (where presumably
we have enough concentration to use densitometric measurements) we can solve
for the initial concentration of target. While this
technique is very attractive, the basic assumption (the equality of
the efficiencies in both species) has some drawbacks \cite{RAEYMAEKERS}. 
Basically,
the potential problems arise in the dependence of the efficiency on the
length of the DNA molecule. The longer molecule will
experience a decrease in efficiency before the shorter one does, as predicted in
Eq.~(\ref{nu}). In any case the model presented here can be of use to assess the
validity of the assumptions that go in the basics of competitive PCR.

In order for competitive PCR to work, the length of the standards 
have to be within a narrow window: it has to be sufficiently different
from the length of the target molecules (to be resolved in a gel) and sufficiently 
similar to it in order for the equal efficiency assumption to work. The
design of a good standard requires some ingenuity, and has to be done on a case
by case basis. In what follows we will present a design for measuring $M_0$
without the need of a standard.
Suppose we measure the concentration of a given DNA molecule after a number 
of PCR cycles on a sample whose $M_0$ is unknown.
One might think that if we repeated the same measurement
for  a reasonable number of times
(say around 100 times, given that PCR equipment with capacity for $96$
vials are not uncommon), so as to  measure the mean value
and the variance of the concentration across that number of experiments, we
would have two equations [Eqs.~(\ref{mean}) and (\ref{variance})] that
can be solved for the two unknowns $p$ and $M_0$. However,
it can be shown that this procedure always yields two possible solutions for 
$p$ and $M_0$,
and there is no possible way {\it a priori}, of choosing the right one. The reason
for this is that for $M_0$ bigger than a few hundreds (which is nonetheless a small
number of molecules), the distribution of $N_n$ is Gaussian, and therefore determined
only by the mean and the variance, which give the above mentioned ambiguous answer.

Consider instead the following scheme. We prepare two sets of samples
$S_1$ and $S_2$,
each with $K$ identical preparations and whose initial concentration of a given
double-stranded DNA molecule is unknown. We run
(under conditions for which $p$
can be considered approximately constant from cycle to cycle)
$n_1$ cycles of PCR on set $S_1$,
and $n_2$ cycles on set $S_2$, after which we measure the number of molecules
in every sample. The averages $\nu_1$ and $\nu_2$ over the $K$ preparations
in  $S_1$ and $S_2$, are estimates of the ensemble averages
$\mu_{n_1}$ and $\mu_{n_2}$ corresponding to Eq.~(\ref{mean}) for $n=n_1$ and $n=n_2$
respectively. We can use that formula to compute
$m_0=\nu_1^{-\frac{n_2}{n_1-n_2}}\nu_2^{\frac{n_1}{n_1-n_2}}$
as an estimate of the
real $M_0$ and $\rho)=\nu_1^{\frac{1}{n_1-n_2}}\nu_2^{-\frac{1}{n_1-n_2}}-1$
as an estimate of the real $p$. Of course these estimates
make sense only if a measure of the error involved in the method is
provided. It takes a simple calculation
to show that,
\be
\la m_0 \ra \approx M_0; \hspace{2cm} \la \rho \ra \approx p \label{mean_estimate},
\ee
and
\bea
\sigma_{m_0} & \equiv & \frac{\la (m_0 -\la m_0 \ra )^2 \ra}{\la m_0 \ra^2}  \approx 
\frac{1}{M_0 K} \frac{1-p}{1+p} \frac{n_1^2 + n_2^2}{(n_1-n_2)^2}, \label{sd_1}\\
\sigma_{\rho} & \equiv & \frac{\la (\rho -\la \rho \ra )^2 \ra}{\la \rho \ra^2}  \approx 
\frac{1}{M_0 K} \frac{1-p^2}{p^2} \frac{2}{(n_1-n_2)^2}.\label{sd_2}
\eea
In writing the last two equations we used Eq.~(\ref{variance}).
We tested these expressions in a set of very simple numerical simulations, whose
details we are not going to report here except for saying that the PCR amplification
was represented by the cascade given by Eq.~(\ref{recursion}). Under variations
of all the parameters involved, Eq.~(\ref{mean_estimate})-(\ref{sd_2}) 
were in excellent agreement with the numerical results. To get a flavor of
the precision of the method proposed, assume a simple example with $M_0=1000$,
$p=0.8$, $n_1=10$, $n_2=15$ and $K=50$. Under these conditions the
above equations predict that
the estimate of $M_0$ will be correct within $0.5 \%$ (that is $\pm 5$ molecules)
and that of $p$ will be correct within $0.1 \%$!. 
These estimates refer to
the purely statistical errors, and they will be fairly small under typical
conditions. In real experiments they have to be supplemented with the errors involved
in the measurement of the concentrations. If $M_0$ and $p$ fluctuated from sample 
to sample (due to inevitable differences
in their preparations), the fact that we are 
averaging over $K$ samples will screen these fluctuations. In this latter case, 
Eqs.~(\ref{mean_estimate}) will still be in agreement with the average
$M_0$ and $p$, and 
Eqs.~(\ref{sd_1}) and (\ref{sd_2}), which can be easily generalized to include these 
fluctuations, will give their right order of magnitude.

\section{SUMMARY}
We have presented a kinetic model for the PCR, which
can be the basis for a more accurate
application of quantitative techniques, as it provides a dynamical
account of the probability of replication as a function of
the physical parameters involved. These include
the rate constants of the different reactions.
Conversely, the model allows us to
extract information on these rates 
from direct measurements of $p$. 
From a theoretical point of view, it 
can also be used in
the description of {\sl in vivo} and {\sl in vitro}
enzymatic polymerization processes \cite{SHAPIRO}.
The statistical analysis of 
PCR under the assumption of constant replication probability shows new 
interesting phenomena. The scaling behavior of the pdf
is an effect of the
recursivity of the process, whereas the
multi-modality is related to failures in
replication during the first cycles. Although the latter is a phenomenon
present only for a small number of initial molecules, it is not far
from actual experimental conditions, and might be of relevance in
quantitative applications.

Finally, we are using the statistical
considerations of section IV to devise 
a method for measuring the initial
number $M_0$ of molecules in a sample (quantitative PCR)\cite{CECCHI}.

\vspace{1cm}

\noindent {\bf Acknowledgements}: Many of the ideas in this paper are the product 
of long and 
fruitful discussions with P. Kaplan and M. Magnasco. We thank
A. Libchaber and E. Mesri for a careful reading of the manuscript and useful
discussions, and an anonymous referee for helpful suggestions.
Support from the Mathers Foundation is 
gratefully acknowledged.

% REFERENCES

\newpage
\noindent {\bf Figure Captions}

\vspace{1cm}
\noindent {\bf Figure 1}. Probability of replication $p(t)$ as a function of
time $t$, for different template lengths $N$ (in number 
of nucleotides without including the
primers), arising from a numerical simulation
of Eqs.~(\ref{chem_reac}),
with parameters:  $\kappa_{1}=
10^{9} M^{-1} s^{-1}$, $\kappa_{2}= 10^{-2} s^{-1}$, $\kappa_{3}= 10^{7}  M^{-1} s^{-1}$, 
$\kappa_{4}=10^{-3} s^{-1}$, $\kappa_{5}= 10^{7}  M^{-1} s^{-1}$, $\kappa_{6}= 15 s^{-1}$, 
$\kappa_{7}=10^{9} M^{-1} s^{-1}$, $\kappa_{8}= 10^{-1} s^{-1}$, $[pr](0)=10^{-6} M$,
$[n](0)=10^{-5} M$, $[q](0)=10^{-6} M$, $[ss](0)=10^{-11} M$.
The square and arrow heads indicate, respectively, the mean rise-time 
and rise-time width as predicted by the simplified model of
Eqs.~(\ref{forward}).

\vspace{0.5cm}
\noindent {\bf Figure 2}. Efficiency $p_{k}$ as a function of the 
cycle number $k$, for different polymerization times. 
The length of the template
is $N=45$; the other parameters are as in Fig. 1.

\vspace{0.5cm}
\noindent {\bf Figure 3}. (a) pdf of the 
number of molecules after $n=10$ cycles and $M_{0}=1$ of a
branching process with
constant efficiency $p$, in log-log scale. 
Notice the multi-modality for $p=0.9$, and the power law
regimes (straight lines). (b) Same as in (a)
for $M_{0}=50$ (linear scale). The multi-modality has
disappeared even for $p=0.99$.

\vspace{0.5cm}
\noindent {\bf Figure 4}. (a) The invariant manifold that crosses
$z=1$ tangent to the unit disk, of the map
 $z_{n+1}=(1-p) z_n + p z_n^2 $ (with $z$ in the complex plane),  for $p=0.9$.
It is parametrized by
$\theta_{\infty,1}(\omega)$. In the same plot the curve parametrized by
$\theta_{15,1}(\omega)$ is shown, and cannot be resolved from
the invariant manifold. (b)
Absolute value of $\theta_{15,1}(\omega)$, and the predicted power
law.
\newpage
\begin{figure}
\label{Figure 1}
\centerline{\epsfysize=10cm \epsfbox{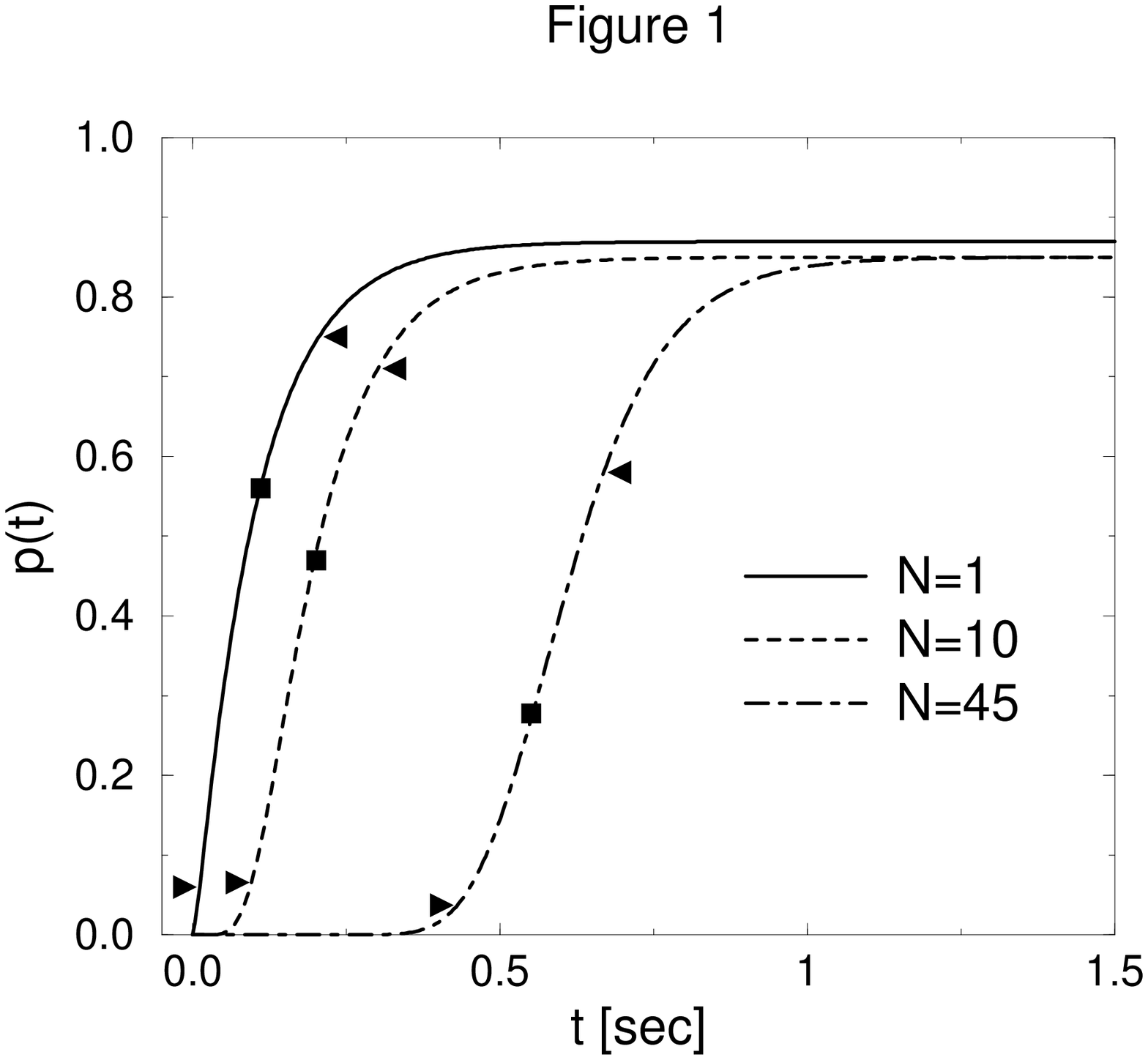} }
\end{figure}

\begin{figure}
\label{Figure 2}
\centerline{\epsfysize=10cm \epsfbox{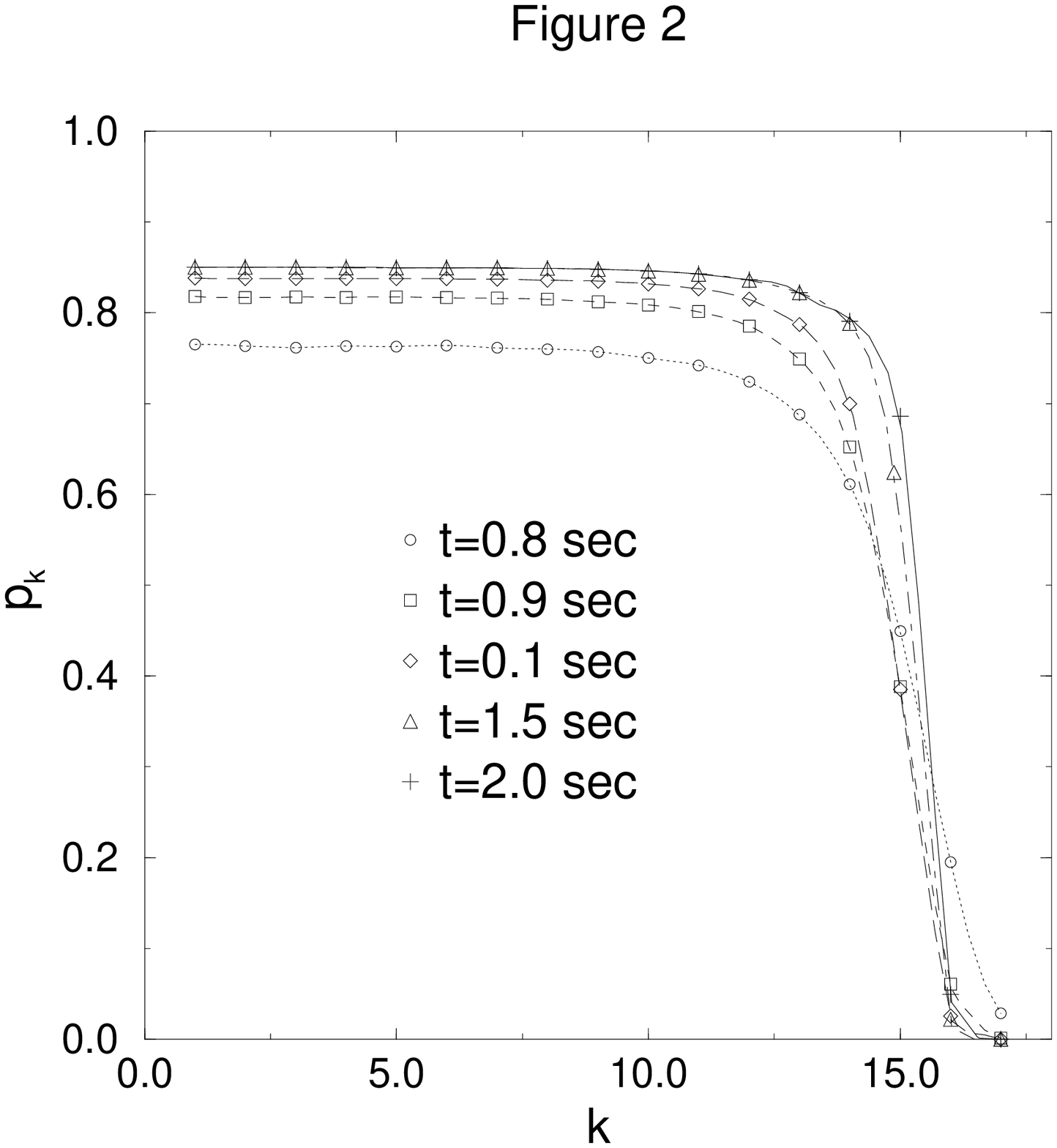} }
\end{figure}

\newpage

\begin{figure}
\label{Figure 3}
\centerline{\epsfxsize=13.2cm  \epsfbox{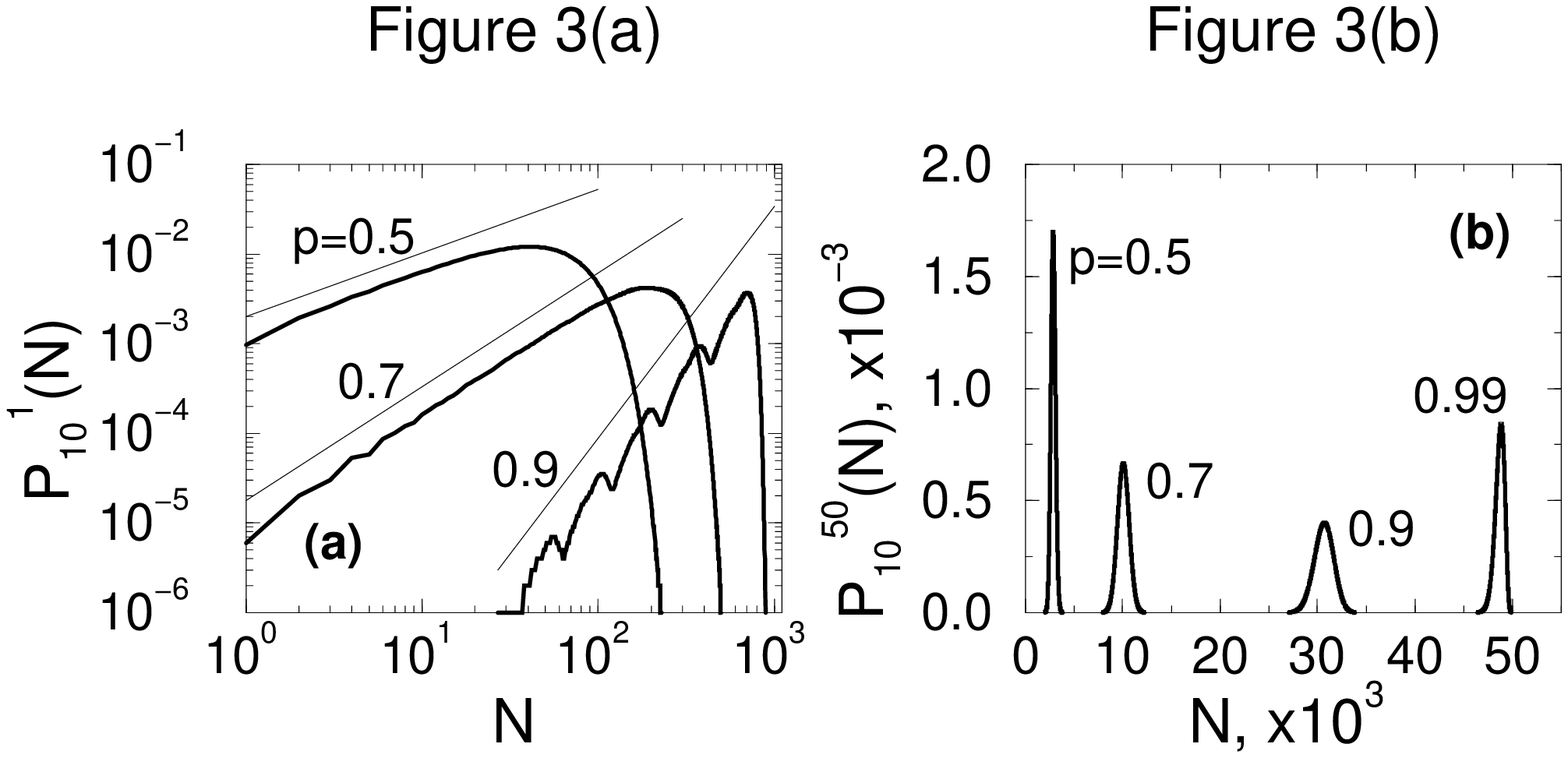} }
\end{figure}

\begin{figure}
\label{Figure 4}
\centerline{\epsfxsize=13.2cm \epsfbox{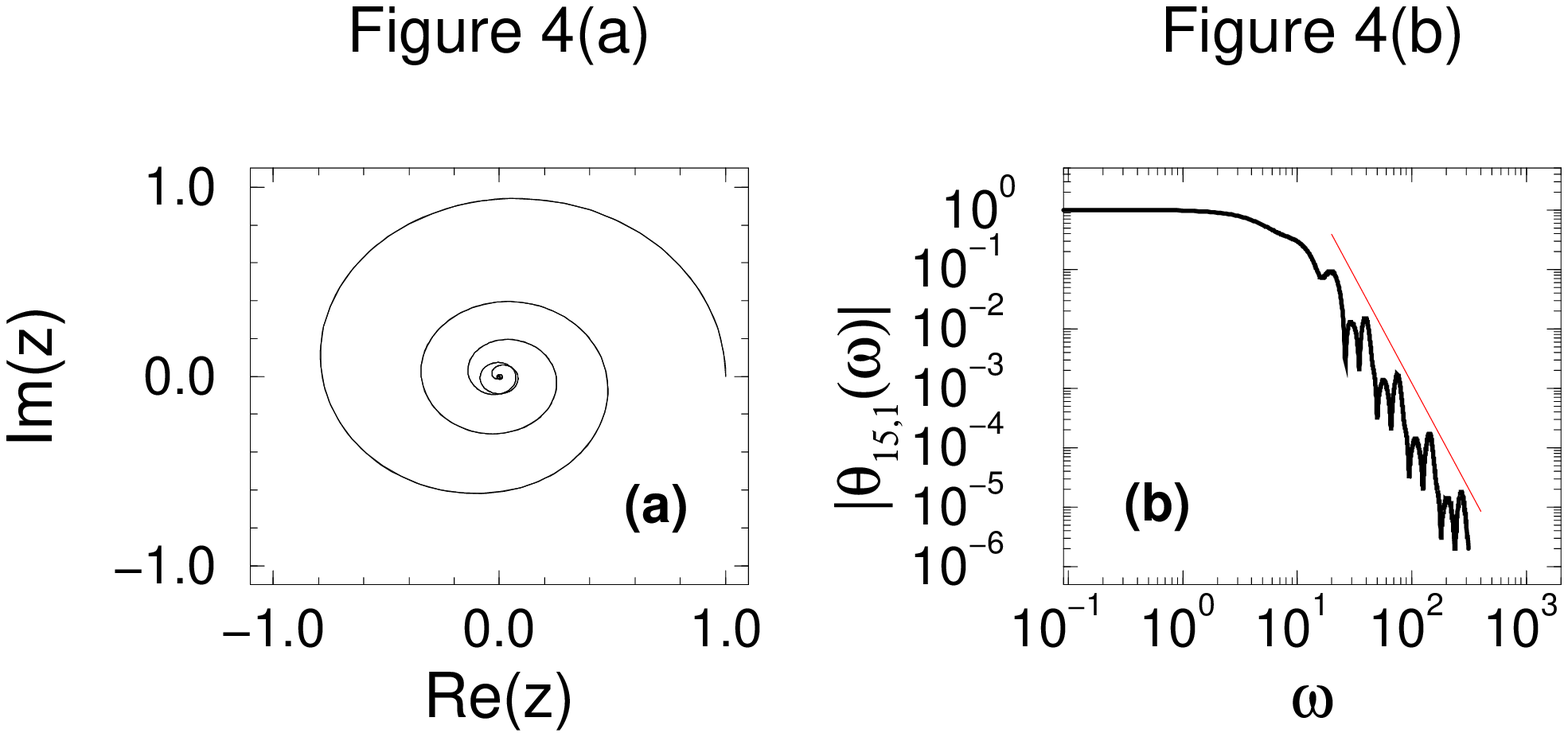} }
\end{figure}

\end{document}